%
%

\input harvmac.tex
\input epsf.tex
\hfuzz 15pt


\def\X{{\goth X}}

\def\Ho{{H_{(1)}}}
\def\Ho{H_{}}
\def\Hoo{H_{(\infty)}}

\def\blank#1{}
\def\bmu{{\mu}}                 
\def\CAL{\CA^<}
\def\comm#1#2{{\big[\;{#1}\;,\,{#2}\;\big]}}
\def\frac#1#2{{{#1}\over{#2}}}
\def\fracs#1#2{{\textstyle{{#1}\over{#2}}}}

\def\sh{{\widehat{sl}}}

\def\vec#1{{|{#1}\rangle}}
\def\cev#1{{\langle{#1}|}}
\def\vac{{\vec 0}}
\def\cav{{\cev 0}}


\def\za{\alpha} \def\zb{\beta} \def\zg{\gamma} \def\zd{\delta}
   
\def\zk{\kappa} \def\zl{\lambda} \def\zm{\mu} \def\zn{\nu}
  \def\zr{\rho}  
 \def\zz{\zeta}

\def\bL{\bar{\Lambda}}
   
\def\zL{\Lambda}  

\def\IZ{Z\!\!\!Z}
\def\[{\,[\!\!\![\,} \def\]{\,]\!\!\!]\,}
\def\dC{C\kern-6.5pt I}

\def\dd{\partial}

\def\bW{\overline W}

\def\CA{{\cal A}}        
        
        \def\CI{{\cal J}}
\def\CJ{{\cal J}}        
    \def\CN{{\cal N}}    
\def\CP{{\cal P}}

\def\un{{\bf 1}}

%

\def\({ \left( }
\def\){ \right) }

\def\dim{{\rm dim\,}}

\catcode`\@=11
\def\Eqalign#1{\null\,\vcenter{\openup\jot\m@th\ialign{
\strut\hfil$\displaystyle{##}$&$\displaystyle{{}##}$\hfil
&&\qquad\strut\hfil$\displaystyle{##}$&$\displaystyle{{}##}$
\hfil\crcr#1\crcr}}\,}   \catcode`\@=12



\def\IC{\relax\hbox{$\inbar\kern-.3em{\rm C}$}}

\input amssym.def
\input amssym.tex
\def\IZ{\Bbb Z}\def\IC{\Bbb C}
\def\gb{\goth b} 


%
%

\def\go{{\goth g}} \def\bgo{\overline{\goth g}} 

\def\bR{\overline\Delta}                   
\def\Rr{\Delta^{\rm re}} 
\def\br{\overline\zr}                      

 

\def\bW{\overline W} \def\tW{\tilde W}      

\def\UNIT{\blacktriangleleft\kern-.6em\blacktriangleright}


\def\ts#1,#2{{\tt e}^{#1\zk#2}}

\def\aa{a}





\def\cc{\chi}        





 
\def\IZpp{\IZ_{>0}}
\def\IQ{{\Bbb Q}}

\def\tri{{\rm tri}}  



\def\la{\langle} \def\ra{\rangle}



\lref\AY{Awata,~H.\ and  Yamada,~Y.,
            Mod.\ Phys.\ Lett.\ {\bf A7} (1992) 1185.}

\lref\BF{ Bernard, D.\ and Felder, G.,
            Commun.\ Math.\ Phys.\ {\bf 127} (1990) 145.}

\lref\D{Dotsenko, Vl. S., Nucl.\ Phys. {\bf B358} (1991) 547.}

\lref\FM{ Feigin, B.L. and Malikov, F.G.,
               Lett.\ Math.\ Phys. {\bf 31} (1994) 315 \semi
          Feigin, B.L.\ and Malikov, F.G.,
               in {\it Operads: Proceedings of Renaissance
               Conferences},   
               Cont. Math. {\bf 202}, p. 357, eds. Loday, J.-L., 
		       Stasheff, J.D. and Voronov, A.A.,  
		       AMS, Providence, Rhode Island, 1997.}

\lref\DLM{Dong. C., Li, H.\ and Mason, G.,
       Commun.\ Math.\ Phys.\ {\bf 184} (1997) 65. }

\lref\K{Kac, V.G.: {\it Infinite-dimensional Lie Algebras}, third
        edition,  Cambridge University Press, Cambridge, 1990.}

\lref\KW{Kac, V.G. and  Wakimoto, M.,
              Proc.\  Natl.\ Sci.\ USA {\bf 85} (1988) 4956
	     \semi 
         Kac, V.G. and  Wakimoto, M.,
               Adv.\ Ser.\ Math.\ Phys. vol {\bf 7}, pp. 138,
               World Scientific, Singapore, 1989
         \semi
         Kac, V.G. and  Wakimoto, M.,
          Acta  Applicandae Math. {\bf 21} (1990) 3.}

\lref\FGPP{
        Furlan, P., Ganchev, A.Ch., Paunov, R. and  Petkova, V.B., 
		       Nucl.\ Phys. {\bf B394} (1993) 665
	       \semi
		   Furlan, P., Ganchev, A.Ch. and Petkova, V.B.,
              Nucl.\ Phys. {\bf B491} [PM] (1997) 635.}

\lref\PRY{ Petersen, J.L., Rasmussen, J. and Yu, M.,
	        Nucl.\  Phys. {\bf B481} [PM] (1996) 577.}		

\lref\FGPboth{Furlan, P., Ganchev, A.Ch. and Petkova, V.B.,
                 Nucl.\ Phys. {\bf B518} [PM] (1998) 645
\semi
           Furlan, P., Ganchev, A.Ch. and Petkova, V.B.,
               Commun.\  Math.\  Phys. {\bf 202} (1999) 701.}

\lref\Zh{ Zhelobenko, D.P., {\it Compact Lie groups and their
               representations}, 
               AMS, Providence, 1973.}

\lref\H{Hiller, H., {\it Geometry of Coxeter groups}, 
               Research Notes in Mathematics {\bf 54},  
		       Pitman Books Ltd, London, 1982.}

\lref\FF{Feigin, B.L. and Fuchs, D.B.,
          J.\ Geom.\ Phys. {\bf 5} (1988) 209.}

\lref\MFF{Malikov, F.G., Feigin, B.L., and Fuks, D.B.,
     	  Funct.\ Anal.\ Pril.\ {\bf 20, 2} (1987) 25.}

\lref\GW{Watts, G.M.T.,
         Commun.\  Math.\  Phys. {\bf  171} (1995) 87.}

\lref\Zhu{Zhu,
         {J.\ Amer.\ Math.\ Soc.} {\bf 9} (1996) 237.}

\lref\AM{Adamovich, D.\ and Milas, A.,
             Math.\ Res.\ Lett.\ {\bf 2} (1995) 563
\semi
Milas, A., {\it Vertex operator algebras associated to modular
        invariant representations}, 
		in {\it `GROUP21: Physical Applications and
        Mathematical Aspects of   Geometry, Groups and Algebras'}, 
		vol. I, p. 167, eds.\ Doebner, H.-D., \ Natterman,
		P. and \ Scherer, W. World Scientific, Singapore, 1997.}
		  

\null
\rightline{ KCL-MTH-99-23}
\rightline{June 1999}
\vskip 2.5cm

\centerline{\bf  
A Note on Decoupling Conditions for Generic Level 
$\widehat{sl}(3)_k$}
\centerline{\bf   and Fusion Rules
}
\footnote{}{ganchev@inrne.bas.bg, ptvp@pt.tu-clausthal.de,
 gmtw@mth.kcl.ac.uk} 

\vskip 1cm
\centerline{  
 A.Ch. Ganchev$^{* }$,        \quad 
 V.B. Petkova$^{* , \dagger}$ \quad 
 and                          ~
 G.M.T. Watts$^{** }$} 
\bigskip
\bigskip

\centerline{\it $^{*}$Institute for Nuclear Research and
Nuclear Energy,} 
\centerline{\it Tzarigradsko Chaussee 72, 1784 Sofia, 
Bulgaria}
\bigskip
\centerline{\it $^{\dagger}$Arnold Sommerfeld Institute for
Mathematical Physics, TU Clausthal,} 
\centerline{\it  Leibnizstr. 10,
38678 Clausthal-Zellerfeld, Germany}
\bigskip
\centerline{\it $^{**}$Mathematics Department,  King's College London,}
\centerline{\it Strand, London WC2R 2LS, UK }
\medskip

\bigskip\bigskip

\centerline{\bf Abstract}

We find the solution of the $\widehat{sl}(3)_k$ singular vector
decoupling equations on $3$--point functions for the particular
case when one of the fields is of weight $w_0\cdot k\Lambda_0$.
The result is a function with non-trivial singularities in the
flag variables, namely a linear combination of ${}_2F_1$
hypergeometric functions. This calculation fills in a gap in
\FGPboth\ and confirms the $\widehat{sl}(3)_k$ fusion rules
determined there both for generic $\kappa \not \in \IQ$ and
fractional levels.

We have also analysed the fusion in $\widehat{sl}(3)_k$ using
algebraic methods generalising those of Feigin and Fuchs and
again find agreement with \FGPboth.  In the process we clarify
some details of previous treatments of the fusion of
$\widehat{sl}(2)_k$ fractional level admissible representations.

\medskip
\medskip
\medskip

%
%
\newsec{Introduction}

A physicist's motivation for studying fractional level WZW 
models is that by Drinfeld--Sokolov reduction one can
obtain a huge class of $W$-algebra models.

Much of the simplicity of integrable WZW models rests in the fact
that all the relevant representations can be induced from finite
dimensional representations of the horizontal Lie algebra --
conversely, all correlation functions can be reduced by the Ward
identities to correlation functions of fields transforming in
finite dimensional representations of the horizontal subalgebra.
Finite dimensional representations can be described in terms of
polynomials on the flag manifold, and so correlation functions of
fields in integrable models also have polynomial dependence on
the flag manifold coordinates.  In 
fractional level WZW models
where the representations are characterised by non integral
weights \KW, the correlation functions of any fields can again be
reduced to functions of flag manifold variables, but these
correlation functions can have singularities not only in the
chiral (`space-time') variable but 
also singularities on
nontrivial submanifolds of the flag manifold.  

The case of $sl(2)$ has been much studied and worked out in some detail.
An example are the
correlation functions of fractional level $\widehat{sl}(2)$ found
in \FGPP, \PRY: the $4$--point blocks are functions of the flag
manifold (``isospin'') coordinate $x$ given by certain multiple
contour integrals and it was shown that there exists a choice of
the contours, depending nontrivially on $x$, such that 
the asymptotic behaviour of the $4$-point blocks reproduces the fusion
rules found by Awata and Yamada (AY) \AY. 
A different fusion rule was proposed somewhat earlier by Bernard and
Felder (BF) \BF, and confirmed by examples of $4$--point correlators
\D{}. The BF rule, however, looks rather degenerate and leads to
nilpotent fusion matrices.   
A more abstract analysis of the number of independent $3$-point
couplings determining the fusion rules was carried out in 
\FM\ and \DLM, who confirmed the fusions rules of AY and BF
respectively, the set of couplings allowed by the BF fusion rule being
a subset of those allowed by the AY rules.
In section 3 we explain clearly the relation between the calculations
of \FM\ and \DLM.

The situation for higher rank cases is much more complicated,
even on the level of $3$--point invariants.  The decoupling of
the Verma module singular vectors is central in any of the
methods used to derive the fusion rules.  The main obstacle comes
from the fact that the expressions \MFF\ for the singular vectors
in Verma modules of non integral highest weights are too
complicated to be analysed as systematically as can be done in
the simplest $sl(2)$ case, or as can be done for the general
integrable representations where (apart from a simple additional
``affine'' vector) the problem reduces to that of the finite
dimensional representations of the horizontal subalgebra.

Despite these difficulties, the fusion rules for the admissible
$\sh(3)_k$ representations at level $\kappa=k+3=3/p$ were found
in \FGPboth\   using only partial information about the solutions
of the decoupling equations.  These rules, which can be easily
extended to the general admissible level values $\kappa=p'/p$,
are expected to be the $sl(3)$ analogue of the generic $sl(2)$ AY
fusion rules.

Moreover, just as the fusion rules at integral $k$ are a
truncation of the ring of tensor products of finite dimensional
representations of $\bgo$, so the fusion rules at a rational
level $\kappa = 3/p$ are truncations of the fusion ring of a
nonrational CFT ($\kappa \not \in \IQ$); similarly, the singular
vector decoupling equations for the representations arising in
this nonrational CFT are a subset of those in the rational CFT.
The fusion ring of this irrational CFT was described in \FGPboth\
where it was explicitly realised by a novel extension of the ring
of formal characters of finite dimensional representations of
$sl(3)$.
It is generated  by a simple current $\gamma\,,$
$\gamma^3=1$ and three `fundamental' representations $f$, $f^*$ and
$h$, satisfying one relation.  
These
representations have (horizontal)  weights $f=-\bL_1\kappa$,
$f^*=-\bL_2\kappa$ and $h=w_{121}\cdot(-(\bL_1+\bL_2)\kappa)$
respectively, where $\bL_1\,, \bL_2$ are the $sl(3)$ fundamental
weights.  The fusion of the fields $f$ and $f^*$ with a generic
field is a sum of seven terms, each with multiplicity one, and in
this way they are analogues of the $sl(3)$ fundamental
representations; the fusion of $h$ however has three terms of
multiplicity one and one of multiplicity two.  In \FGPboth, the
singular vector decoupling equations were used to examine the
space of three point couplings, and the (generically) $7$-terms
fusion of $f$ was found, leading also to explicit solutions
for the $3$--point invariants.  However, for fields which
admitted fusions with non-trivial multiplicities, the number of
three-point couplings with a generic field found in \FGPboth\ was
less than that predicted by the fusion ring.

In this paper we reconsider this problem and analyse the
decoupling conditions for the simplest nontrivial multiplicity
examples by two different  methods.  In section 2 we use 
(as in \FGPboth) the standard method of representing the generators
and the singular vectors in terms of differential operators with
respect to the flag variable coordinates $X$.  
Whereas in \FGPboth\ an explicit ansatz was used for the three-point
coupling, here we consider an arbitrary function of $X$;
we find that for the representation $h$ the null-vector decoupling
conditions reduce to two hypergeometric equations. For the
multiplicity two state we get two solutions spanned by appropriate
${}_2F_1$ hypergeometric functions. Similar analysis applied to a more
complicated example involving $4$-order differential equations also
confirms the fusion rule  multiplicities  of \FGPboth. 
 
The second method we exploit in section 3 is purely algebraic,
extending a method of  Feigin and Fuchs \FF\ for the Virasoro
algebra, and of Malikov and Feigin \FM\ for $\hat{sl}(2)$.
It is based on the fact that, while the fields $\Phi(X)$
transform under a differential operator realisation of
$\bgo=sl(3)$ which is neither highest nor lowest weight
representation, these fields can be also interpreted as highest
weight states with respect to a choice of the Borel subalgebra
$\bar{b}(X)$ depending on the coordinates $X$ of the flag
manifold.  The idea is to consider the space of states generated
by the first field in a $3$--point function and to use  the fact
that the two other fields are highest/lowest weight states after
an appropriate rotation of the Borel subalgebra. Thus this space
of states is effectively factored by an ideal depending on $X$
and the truncation is so severe that we are left with a finite
dimensional space describing the possible fusion rules.  In
section 3 we start first with the simpler case of $sl(2)$ where
we discuss the relation between the fusion rules in \AY\ and
those in \BF. In the $sl(3)$ case, the algebraic approach is
applied to the couplings of the representation $h$ for generic
values of the coordinate $X$. The two approaches to the
decoupling problem are in full agreement and  confirm the fusion
rule of \FGPboth.

\medskip


\subsec{Fields and generators}

In this paper we shall deal with representations of
$\go=\widehat{sl}(3)_k$ labelled by non-integral weights.  When
(as is typically the case) the horizontal projection of the
non-integral weight is also non-integral, the associated field
does not transform in a finite dimensional representation of the
horizontal subalgebra $\bgo$.

The infinite dimensional representations which arise this way are
described classically by induced representations of $SL(3)$
on functions $f: SL(3) \mapsto \IC$ satisfying the condition
$f(g'YH)=\chi_{\mu}(H) f(g')$, where
$\chi_{\mu}(H)$ is a character of the Cartan subgroup and $Y$ is the
subgroup of upper triangular matrices with units on the diagonal. 
The representation is given by the left multiplication of $SL(3)$
on the space of functions, $(T(g)f)(g')= f(g^{-1} g')$.

Equivalently these representations are realised in terms of
functions on the homogeneous $G/(YH)$ space obtained by dividing
by the Borel subgroup $YH$. This space is the flag manifold
$\cup_{w\in \bW} \, \hat{w}\, {\X}_w$, where $\bW$ is the Weyl
group with elements $w$ represented  by matrices $\hat{w}\in
SL(3)$.  The generic  3-dimensional ``big cell''
${\X}={\X}_{\un}$  consists of lower triangular matrices with
units on the diagonal; we shall denote them  $X(x,y,z)$, or,
simply $X$, where $(x,y,z)$  are coordinates  corresponding to
the matrix entries $21$, $32$ and $31$ respectively (see, e.g.,
\Zh,  or, for a broader discussion, \H).  The remaining cells
$\hat{w}\, {\X}_w$, with $ {\X}_w$ given by lower triangular
matrices with  some zeros, account for the ``infinities''
encountered in the (local) Gauss decomposition  of the elements
of $SL(3)$ and are thus needed to give meaning to the global left
action of the group.  Hence these remaining cells can also be
thought of as the result of taking some of the coordinates in $X$
to infinity in certain prescribed ways.  In particular,
$\hat{w}_{\theta}=\hat{w}_{\theta} X(0,0,0)\,$ is the only point
of the cell of lowest dimension of the flag manifold where
$\hat{w}_{\theta}$ is the  matrix realising the action of the
longest Weyl group element $w_{121}{\,=\,}w_{\theta}$, coinciding
in the $sl(3)$ case with the  reflection with respect to the
highest root $\theta{\,=\,}\za_1{+}\za_2$.

The right action of the group, $(T^R(g)f)(g') = f(g'\,g)$, or
rather its  infinitesimal version, has a different meaning:
because of the above invariance with respect to $YH$, any of the
functions $f$ can be identified with a highest weight state of a
highest weight representation  built by the generators of (right)
translations. 

Following the classical analogy, the operator valued quantum
fields (chiral vertex operators) depend on a pair of variables,
$\Phi_{\zm}(X)\equiv\Phi_{\zm}(X,u)$, where $u$ is the usual
space-time chiral coordinate. The complex number $u$ can be
similarly interpreted as a coordinate on the big cell of the flag
manifold of $SL(2)$,  the action of the non-trivial Weyl group
element $r$ is given by  $\hat r(u) {\,=\,}{-}1/u$, and the point
at infinity on the Riemann sphere is the only point $\hat r(0)$
on $\hat{r}\,{\X}_r$.

The commutation relations of the generators $T_n$ of $\go$
with these fields are
\eqn\aa{
[\Phi_{\zm}(X,u)\, ,\, T_n]
=u^n\,D^{(\mu)}(T)\, \Phi_{\zm}(X,u)
}
where $T\equiv T_0\in\bar\go $ and $D^{(\mu)}(T)$ are differential
operators (the infinitesimal version of the above left action of the
group). In our convention these are
\eqn\do{\eqalign{
D(f^1)& = - \dd_x,\quad  
D(f^2) = - (\dd_y +x \dd_z),\quad 
D(f^3) = - \dd_z\,, \cr
D(h^1)& = 2 x \dd_x-r_1 -y \dd_y + z \dd_z\,, \cr  
D(h^2)& = 2 y \dd_y-r_2 -x \dd_x + z \dd_z\,, \cr
D(e^1)& =  (x)^2 \dd_x-r_1 x + (z-x y) \dd_y + x z \dd_z\,,\cr  
D(e^2)& =  (y)^2 \dd_y-r_2 y -z \dd_x\,,\cr
D(e^3)& =  (z)^2 \dd_z-r_2 (z-x y)-r_1 z+ x z \dd_x 
 +y (z-x y) \dd_y \,,
}}
where for simplicity we have omitted the index ${\mu}$.  Here
$r_i{\,=\,}\la \mu\,, \alpha_i\ra{\,=\,}\mu(h^i) $ are the
components of the weight $\mu$.  This is  a
representation of $\go$  with $k=0$.

One can define states $|\nu\ra=\lim_{ x,y,z\,, u\to 0}
\Phi_{\nu}(X,u)|0\ra$, which satisfy the usual conditions of a
highest weight state
\eqn\hws{
e^i|\nu\ra=0\,, \quad 
h^i|\nu\ra=- D^{(\nu)}(h^i)|\nu\ra=\nu(h^i)\,|\nu\ra\,,\quad 
T_n \, |\nu\ra=0\,, \ n>0\,,
}
where furthermore $|\nu\ra$ is assumed to be an eigenstate of the
central charge operator with eigenvalue $k$; we shall omit the
explicit dependence on $k$ of this highest weight state.
Analogously, the dual state
$\cev\zg$ satisfying $\cev\zg\nu\ra=\zd_{\nu\zg}$ and
\eqn\dhw{
  \la\zg| f^i = 0 
\,,  \quad
  \la\zg| h^i = \zg(h^i) \la\zg|
\,,\quad 
  \la\zg| T_{-n} =0\,,\;\; n>0\,.
}
can be reproduced, with a proper normalisation of the fields, as  
\eqnn\dual 
$$\eqalignno{
   \la\zg|
&= \lim_{x,y,z\,,u\to 0} 
   \cav\,
   \Phi_{\zg^*}(\,\hat{w}_{\theta}X(x,y,z), \hat{r}(u)\,)
&\dual \cr
&= \lim_{x,y,z\,,  \,u\to 0}\, 
   z^{\zg(h^2)} (z-xy)^{\zg(h^1)}\, u^{-2\triangle_{\zg}} \,
   \cav\,
   \Phi_{\zg^*}\big(\,X\big({x\over z}\,,\, 
                    {y\over xy{\,-\,}z}\,,\, 
                    {1\over z}\big)\,,\, 
                -{1\over u}\big) 
\;,
}$$
where $\Delta_{\zg}$ is the Sugawara conformal weight and
$\zg^*{\,=\,}{-}w_{\theta}(\zg) $.  Note that this way the limits
of the chiral coordinate $u$ and the flag coordinate $X$ are
treated on the same footing.  Eqn.\ \dhw\ follows from \dual\
using
$$
  \comm{\Phi_{\zg}(\,\hat{w}_{\theta}X, \hat{r}(u)\,)}
       {T_n}
= (-u)^{-n}\,D^{(\zg)}(\hat{w}_{\theta}(T)) \, 
  \Phi_{\zg}(\,\hat{w}_{\theta}X, \hat{r}(u)\,)
\;,
$$
where $\hat{w}_{\theta}(T)$ is the adjoint action of
$\hat{w}_{\theta}$ on $T$.

In section 4 we discuss a purely algebraic treatment of
correlation functions which relies on  the use of a set of
generators which can be seen as the analogue of the right action
generators discussed above. Namely we define $\hat{T}_n\equiv
T_n(X)$ by
\eqn\rac{
  \hat{T}_n 
= U(X)\,T_n \, U(X)^{-1}
\,,
}
where  $U(X)=\exp(xf^1+yf^2+(z-{xy\over 2})\,f^3)$ is the 
operator implementing the translations
$$
  U(X')\, \Phi_{\mu}(X)\, U(X')^{-1}
= \Phi_{\mu}(X'X)
\,.
$$
(The corresponding infinitesimal generators appear in the
first line of \do.) More explicitly \rac\ read
\eqn\race{\eqalign{
\hat{f}^1&= f^1 + y  f^3\,, \qquad 
\hat{f}^2 = f^2-x  f^3\,, \qquad
\hat{f}^3 =  f^3\,,\cr
\hat{h}^1&= h^1 +2 x f^1 - y f^2 +(z{+}xy) f^3\,,\cr
\hat{h}^2&= h^2 +2 x f^2 - y f^1+(z{-}2xy) f^3\,,\cr
\hat{e}^1&= e^1 -x h^1-x^2 f^1 +z f^2 - xz  f^3\,,\cr
\hat{e}^2&= e^2 -y h^2- y^2 f^2 - (z{-}xy) f^1- y(z{-}xy) f^3\,,\cr
\hat{e}^3& =e^3 -  (z{-}xy) h^1 -z h^2 -y e^1 +x e^2 -x (z{-}xy) f^1
-y z f^2 -z(z-xy) f^3\,.
}}

If the generators in the r.h.s.\ of \race\ are replaced by their
differential operators  counterparts in \do\ then they reduce (up to a
sign) to the classical generators of the right action of the group,
i.e.\ $\hat{f}^i$ turn into the generators of right translations,
$\hat{e}^i$ vanish identically while $\hat{h}^i$ reduce to the
numerical values $-\zm(h^i)$. Thus \rac\ can be also looked on as a
change of basis in the algebra which partially diagonalises the left
action of $\go$. (Such an algebra $\go(X)$ attached to a point $X$
of the flag manifold has been discussed in the $sl(2)$ case in \FM.)
Explicitly, choosing $u=1$
\eqn\Ia{\eqalign{
 [\Phi_{\mu}(X)\,, \,  \hat{h}_n^i]&=-\zm(h^i)\,
\Phi_{\mu}(X)\,,\cr 
 [\Phi_{\mu}(X)\,, \, \hat{e}_n^j]&=0\,.
}}

\medskip
\subsec{Correlation functions and singular vectors}

One can now address the problem of finding $n$--point functions
invariant with respect to the action of $\bgo$ in \aa, \do.
While the  full  $3$--point correlators 
$$
  \cav\,
  \Phi_{\zg^*}(X_3,u_3)\,
  \Phi_{\zm}(X_2,u_2)\,
  \Phi_{\nu}(X_1,u_1)\,
  \vac
\;,
$$
are indeed invariant w.r.t.\ $\bgo$,
it is simpler to deal with the correlator with fixed first 
and third coordinates
\eqn\cpl{
C_{\zm \zn}^{\zg}(X)=\la \zg| \Phi_{\zm}(X) |\nu \ra
 \,,
}
which are only restricted by the counterpart of the Ward identity
with  respect to the Cartan generators, 
\eqn\wi{ \Big(\nu(h^i)- \zg(h^i) - D^{(\zm)}(h^i)\Big)  C_{\zm
\zn}^{\zg}(X)=0\,, \quad i=1,2\,.
}
Similarly the Ward identity with respect to the scale generator
$L_0$ fixes the dependence on $u$ (suppressed in \cpl{}) to a
power given by $\triangle_{\zg}-\triangle_{\nu}-\triangle_{\mu}$.

The requirement that singular vectors in the Verma module of
highest weight $\nu$ decouple from correlation functions imposes
additional restrictions on \cpl, thus selecting a subset of the
possible $3$--point invariants.  The dimension of the total
(linear) space of solutions gives the multiplicity $N_{\mu
\nu}^{\zg}$  of the representation $\zg$ occurring  in the fusion
$\zm\otimes \zn$.

There is a singular vector (denoted by $S_\beta\vec\nu$)
in the Verma module with highest weight $\nu$ 
whenever there is a real positive root $\zb \in 
\Rr_+=\bR_+\ \cup\ (\bR+\IZpp\ \zd)$
satisfying the Kac-Kazhdan reducibility condition 
\eqn\KK{
  \la \nu+ \br+ \kappa \zL_0\,, \, \zb \ra \in \IZ_{>0} \;.
}
Here $\zd{\,=\,}\za_0{\,+\,}\za_1{\,+\,}\za_2$; $\alpha_j\,,
j{=}0,1,2\,,$ are the three simple roots of $\go$; 
$\br=\bL_1{+}\bL_2$
is the Weyl vector of $\bgo$; $\kappa=k+3\,$ is the shifted value of
the central charge; and $\zL_0$ is the fundamental weight of $\go$
dual to the affine root $\za_0$ satisfying 
$\la \zL_0,\alpha_j\ra= \zd_{j0}$.  (for details see, e.g., \K{}).
$S_\beta\vec\nu$ is a highest weight vector with weight
$w_\beta\cdot\nu$ given by the shifted action of the Kac-Kazhdan
reflection $w_{\zb}$ in the affine Weyl group $W$ on $\nu$.

The element $S_\beta$ is of a  fixed grade $n<0$ in the
universal envelope of $\go_-$  ($\go=\go_-\oplus\go_0\oplus\go_+$
being the triangular decomposition  of $\go$). Using this fact,
eqns.\ \aa\ and \dhw, commuting $S_\beta$ through the field we find
\eqn\dezero{\eqalign{
  \cev\zg\,
  \Phi_\mu(X,u)\,
  S_\beta\,\vec\nu
&= u^{-n}\,
  D^{(\zm)}(\,\bar S_{\zb}\,)\,
  \cev\zg\,
  \Phi_\mu(X,u)\,
  \vec\nu
\;,
}}
where $\bar S_\beta$ is the projection of $S_\beta$ to the
horizontal subalgebra.  Imposing the vanishing of the matrix
elements of $S_\beta\vec\nu$ leads to the differential equation
\eqn\de{
  D^{(\zm)}(\,\bar S_{\zb}\,)\  C_{\zm \zn}^{\zg}(X)=0\,.
}
We turn to the solutions of these equations in section 2.

\subsec{ The model}

We now turn to our problem in which we shall essentially deal
with a non-rational counterpart of the admissible CFT \KW.

We shall call ``pre-admissible'' the infinite set ${\cal P}_+$ of
highest weights defined for generic ($k \not \in \IQ$) values of
the central charge  considered in \FGPboth,
\eqn\pre{ 
  \CP_+
= \{ \zL\equiv wt_{-\zl}\cdot(\lambda' + k\zL_0) \,|\, 
     w\in \bW\,,\;\;  
     \zl, \zl' \in P_+\,,\;\;
     \la\zl,\za_i\ra \delta +w(\za_i)\in \Rr_+\,,\;\;  i=1,2 
  \}\,.
}
Here   $t_P$ is the group of translations in the weight lattice
$P=\oplus_{i}\ \IZ\ \bL_i$  of the horizontal subalgebra $\bgo$.  
For such weights the Kac-Kazhdan roots $\beta_i$, $i{=}1,2$ are explicitly
\eqn\preKK{
     \zb_i=\la\zl,\za_i\ra \delta +w(\za_i)
\;,
}
and the reducibility pattern of the $\go$ Verma modules with
highest weights in this  set parallels that of the Verma modules
of $\bgo$ of dominant integral highest weights; in particular,
the maximal submodule of any such $\go$ Verma module is a union
of the Verma submodules generated by the two singular vectors
$S_{\zb_i}|\zL\ra$ with $\zb_i$ as in \preKK.  In what follows we
shall mostly use the horizontal projections 
$\bL=w\cdot(\zl'-\zl \kappa)$  of the  weights $\zL\,.$

If in \pre\ we take $w{\,=\,}\un\,, \zl{\,=\,}0$, we recover the
weights of a non-rational CFT (which might be called
``pre-integrable'')  for which the fusion rules are given by the
standard $sl(3)$  tensor product rules for $\lambda'$.

Here (as in \FGPboth) we shall mostly concentrate on the other
generic subset of \pre\ obtained by taking $\zl'{\,=\,}0$.  The
set of such weights can be looked on as the set of ``integral
dominant'' weights  of the ``weight lattice'' $\tW=\bW\ltimes
t_P\,$ -- the extended affine Weyl group of $\go$.  
Each of these weights has associated to it a generalised weight diagram
parametrised by a finite subset of the affine Weyl group
$W{\,=\,}\bW\ltimes t_Q\,$ (where  $Q$ is the root lattice of
$\bgo$), and a generalised formal character $\cc_{_{\Lambda}}$.
The characters close under multiplication and the structure
constants of the resulting (commutative) ring serve as the fusion
rule multiplicities of the corresponding non-rational CFT.  The
ring is an extension of the ring of characters of finite
dimensional representations of $sl(3)$.  It is generated by three
``fundamental'' characters and a ``simple current'' character
with highest weights  given by
\eqn\fun{
 f= -\bL_1\kappa\,, \quad   f^* = -\bL_2\kappa\,, \quad 
 h=\overline{w_0\cdot (k\Lambda_0)}
  = w_{121}\cdot (-\bar{\rho} \kappa)
  =\bar{\rho}(\kappa-2)\,,
 \quad  \gamma = w_{12}\cdot(-\bL_2\kappa)  \;,
}
and which satisfy 
\eqn\const{
 \cc_{_{\gamma}}^3 = 1\;,\qquad\quad
 \cc_{_{ h}}\,  \cc_{_{ h}}  = 2\,  \cc_{_{h}} \;+\; \un\;+
  \cc_{_{\gamma}}\, \cc_{_{f}} \;+\;  \cc_{_{\gamma^2}}\,
  \cc_{_{f^*}}\;.
}
The singular vectors in the Verma modules of highest weights
\fun\ are determined by the corresponding Kac-Kazhdan reflections 
$w_{\zb_i}\,, i=1,2$ and can be recovered from the general
formulae of \MFF.  For the simple current $\gamma$ these are
especially simple $(\zb_1\,,\zb_2)=(\za_2\,, \zd-\za_1-\za_2)$,
i.e.\ the two singular vectors are given by monomials of $f^2$
and $e_{-1}^3$, and the rules for the fusion of $\zg$ with an
arbitrary weight $\bmu$ are correspondingly simple: 
$\zg\otimes \bmu =w_{12}\cdot(-\bL_2\kappa +\bmu)$.  The fusion rules
and the generalised formal  characters of the remaining
representations in \fun\ are described in \FGPboth. In particular the 
representation denoted $h$ provides the simplest example of a
nontrivial multiplicity and its fusion with a generic $\bmu$ reads
\eqn\fush{
  h\otimes \bmu 
= 2 \bmu \,
  \oplus\,
  w_{121}\cdot(-\br\kappa + \bmu) \,
  \oplus\, 
  w_1\cdot(\bmu) \, 
  \oplus\,
  w_2\cdot(\bmu)\,.
}
In the next section we shall recover this generic fusion rule
from the decoupling of the singular vectors in the Verma module
of highest weight $h=\bar{\rho}(\kappa-2)$.

\newsec{Differential equation approach}

We start by recalling the result in \FGPboth\ for the first
two of the fundamental representations in \fun.  In \FGPboth\ the
decoupling equations corresponding to the real positive roots
$\zb_i \,, \, i=1,2$ were investigated using the ansatz
\eqn\monos{
  x^a (z-xy)^c y^b z^d \;.
}
We recall that the solution of \de\ in the integrable case is
given by such monomials with nonnegative integer powers $a,b,c,d$
restricted by $0\le a+c\le \la \nu, \za_1 \ra\,, \ \ 0\le b+d\le
\la \nu, \za_2 \ra\,, $ a basis  being selected, e.g., by the
subset $\{(a,0,c,d)\}\cup\{(0,b,c,d), b\not=0\}$  (see
e.g.,\Zh{}).  Choosing $\zm$ (or $\zg$) to be the identity
representation, these monomials and the corresponding values of
$\zg=\nu -a\alpha_1 -b\alpha_2 -(c+d)(\za_1+\za_2)$ (or $\mu$)
are in one to one correspondence with the weight diagram of the
finite dimensional representation of highest weight $\nu$ (or
$\nu^*$ respectively) and the number of different monomials
producing a given value of $\zg$ coincides with the multiplicity
of this weight in the weight diagram.
			
The solutions found  in \FGPboth\ involved monomials \monos\
with nonintegral powers.  However it is not necessary to assume
such an ansatz which is too restrictive in general
and we present here an alternative derivation.

Since the ratio $\zeta=z/(xy)$ is invariant with respect to the
global $SL(3)$ scale transformations $(x,y,z)\rightarrow
(\zr_1x\,,\,\zr_2y\,,\, \zr_1 \zr_2 z)$, (i.e. it is annihilated
by the Cartan generators $D^{(\mu)}(h^i)$ in \do\ for
$r_i=\mu(h^i)=0$) the general solution to the Cartan subalgebra
Ward identities \wi\ is
\eqn\tri{
C(x,y,z)=x^A y^B\, F\big({z\over xy}\big)\,,
}
where  $F(\zeta)$ is an arbitrary function and $A$ and $B$ are
determined by \wi:
\eqn\wia{\eqalign{
  \la\nu{\,+\,}\zm
- A\za_1 - B\za_2 - \zg\,, \,   \za_i\ra=0\,, \quad i=1,2 \,.
}}
When $\nu$ is one of the analogues of the symmetric
representations $\nu= -l\bL_1 \kappa\,$ or $\nu= -l\bL_2
\kappa\,$ one of the differential operators is very simple --
$D(f^2)$ or $D(f^1)$. This restricts $F(\zz)$ to a monomial
$F(\zz) = (\zz{-}1)^B$ or $F(\zz) = (\zz)^A$, respectively, and
hence the solutions are indeed monomials of the form \monos.  It
then remains to use the equation corresponding to the other
singular vector to determine the possible values of $\zg$ as a
function of $\nu\,, \, \zm$.

This has been done in \FGPboth\ for $l=1$ and $l=2$.  For the
representation $f$ given by \fun\ there turn out to be $7$
solutions for $\zg\in {\cal P}_+ \,$ in \wi, i.e. $7$ terms in
the fusion  $f\otimes \mu$ for generic values of $\mu\in {\cal
P}_+ \,.$ For completeness we write down here the zero mode
projection of the nontrivial  singular vector for $l=1$, i.e.
for the first example in \fun\
$$
\eqalign{
 \bar S_{\zb_1}
&= (f^1)^{1{+}\kappa}\, 
   (e^3)^{2{-}\kappa}\, 
   f^2\,
   (e^3)^{\kappa{-}1}\, 
   (f^1)^{1{-}\kappa} \cr
&= \big(   e^3\, f^1+(1{+}\kappa)\,e^2\big)
   \big(   f^2\, f^1 {-} \kappa  \,f^3\big)
+ (1{-}\kappa)\,
  \big(e^1\,f^1 - (1{+}\kappa)\, h^1
-\kappa\,(1{+}\kappa)\big)\,f^1   \,.
}
$$

We turn now to the third ``fundamental''  weight $h$ in \fun, for
which the two singular vectors correspond to the roots $\zb_i=
\za_0+\za_i\,, \ i=1,2\,. $ Their horizontal projections are
\eqn\sva{\eqalign{
   \bar S_{\zb_1}
&= e^3\, f^1 
   + (2{-}\kappa)\, e^2 \,, \cr
\bar S_{\zb_2}
&= e^3\, f^2 - (2{-}\kappa)\, e^1 \,.
}}
Inserting the differential operator representation \do\
one obtains two second order differential equations for the
unknown function $F$ in \tri\
\eqnn\hyp $$\eqalignno{
& \Big[ \;
  \zeta^2 ( 1  {-} \zeta ) {d^2\over {d\zeta}^2}
 \,-\,
  \zeta\Big((1{+}A{+}B{-}\kappa{-}r_2)
 + \zeta(\kappa{-}2 A{-}B{+} r_1 {+}r_2)\Big) {d\over d\zeta}
\cr
&+
  \Big(    (\kappa{-}2{-}A)(r_2{-}B) 
       -  A\zeta(1{+}A{+}B{-}\kappa{-}r_2{-}r_1)
  \Big) 
  \;
  \Big] F(\zeta)=0\,, 
& \hyp \cr
& \Big[
  \;
  \zeta (1{-}\zeta)^2  {d^2\over {d\zeta}^2}
 \,-\,
 \Big( B{-}1{-}r_2{+}\zeta(1{-}A{-}3 B{+}r_1{+}2 r_2{+}\kappa) 
 + {\zeta}^2 (A{+}2 B{-}r_1{-}r_2{-}\kappa )
  \Big){d\over d\zeta} \cr
&-\Big( B (1{+}B{-}r_2{-}\kappa) 
        {+}(A{-}r_1)(\kappa{-}2) 
       -B\zeta(1{+}A{+}B{-}\kappa{-}r_2{-}r_1)
   \Big)
   \; 
   \Big] F(\zeta)=0\,. 
}$$
For generic values of $A,B$ the two equations in \hyp\ are
different and their order can be reduced exploiting Euclid's
algorithm -- for generic values of $\mu$ the second order terms
can be eliminated between the two equations giving a first order
equation; differentiating this equation again we can then
eliminate the second order term from one of the two original
equations -- this whole process yielding two first order
differential equations.  For generic values of $\mu$ the
resulting system of two first order equations is consistent for
three particular values of the pair of parameters,
$(A,B)=(r_1+r_2, r_1+r_2)\,, \, (r_1+\kappa-1,\kappa-2)\,,\,
(\kappa-2,r_2+\kappa-1)$, and accordingly one obtains three
monomial solutions $F(\zeta)=(\zeta-1)^c
\zeta^{d}$ with $(c,d)=(r_2, r_1)\,, \, (0,\kappa-2-r_2)\,,
\,(\kappa-2-r_1, r_1)$, respectively. Inserting the values of
$(A,B)$ in \wia\ we obtain for a given generic $\mu$ three values
of $\zg$ which precisely recover the remaining multiplicity $1$
weights in the fusion $h\otimes \mu$ as predicted in \FGPboth.

However, for $A{\,=\,}B{\,=\,}\kappa{-}2$ the two equations \hyp\
become identical, and the resulting equation is a  hypergeometric
equation.  For such $A,B$ and $r_2\not=\kappa{-}2$ the solution
of \hyp\ is a linear combination of  two hypergeometric functions
\eqn\thy{
\eqalign{
   F_1(\zeta)
&= {_2}F_1(2-\kappa\,,\,3+r_1+r_2-\kappa;\,3+r_2-\kappa;\,
\zeta)\,, \cr 
   F_2(\zeta)
&= (\zeta)^{\kappa-2-r_2}\, {_2}F_1(-r_2\,,\,1+r_1;\,
   \kappa-1-r_2;\, \zeta)\,. 
}}
For $r_2 {\,=\,}\kappa{-}2$ these hypergeometric series
formally coincide so one of the solutions is instead logarithmic,
\eqn\ln{\eqalign{
   F_1(\zeta)
&= {_2}F_1(2-\kappa\,,\,1+r_1;\,1;\, \zeta)\,,\cr
   F_2(\zeta)
&= {\rm ln}(\zeta)\ {_2}F_1(2-\kappa\,,\,1+r_1;\,1;\,
\zeta)\, + \, \ldots
}}

To find the representation $\zg$ to which this fusion corresponds
we  insert $A=B=\kappa-2$ in  \wia\ to find $\zg=\zm$, i.e. the
weight $\zm$ appears twice in the fusion $h\otimes
\zm$. This is the missing multiplicity two solution predicted
in \FGPboth\ which is now seen to correspond to a novel
hypergeometric function expression of the matrix element $C(X)$
\tri. In fact a multiplicity $2$ was noticed in the previous
calculations but only for the particular choice $\mu=0$.  We can
reconcile this with the treatment here by noticing that for
$r_1=0$ the first hypergeometric series in \thy\ simplifies to a
geometric series, i.e,
$F_1(\zeta)=(1-\zeta)^{\kappa-2}=(xy)^{2-\kappa}\,
(xy-z)^{\kappa-2}$, and similarly for $r_1=r_2=0$ the second
solution in \thy\  becomes
$F_2(\zeta)=\zeta^{\kappa-2}=(xy)^{2-\kappa}\, z^{\kappa-2}$,
i.e.\ both solutions degenerate to  monomial solutions of the
type discussed above.

As another example we have checked by Mathematica the case with
$\nu=-\br \kappa$, the analogue of the adjoint representation of
$sl(3)$. It involves a system of two  fourth order linear
differential equations for the function $F(\zeta)$ and reveals a
relation between the order of the equation and the multiplicity
of the corresponding value of $\zg$ similar to the one
encountered in the above simpler example. 
Namely for the particular values of the parameters 
$A=B=-\kappa$ 
the initial system degenerates to one fourth order differential  equation
possessing $4$ linearly independent solutions and thus leading to
a multiplicity $4$ of the weight $\zg=\zm$;
If $\gamma\neq\mu$ these two fourth-order equations can be reduced to
two third order equations but there are three different sets of values
of $A$ and $B$ for which these third order equations become identical, 
leading to three solutions of multiplicity three;
at the 
next step there are 9 possibilities of degeneration to one second
order equation (again a hypergeometric equation) leading to
multiplicity $2$ solutions, and finally there are $12$ monomial
solutions of first order equations corresponding to multiplicity
$1$ values of $\zg$. The final result is in perfect agreement
with the prediction in \FGPboth.

One can expect that a similar mechanism holds in general. In
particular the maximal order of the differential operator
corresponding to a singular vector $S_{\zb_i}$  (with $\zb_i$ as
in \preKK) can be computed from the expression in \MFF\ to be
$\la 3\zl+w(\br)\,, \, \za_i\ra$.  The minimal of these two
numbers coincides precisely with the maximal multiplicity in the
generalised weight diagram associated to $\bL=w\cdot(-\zl
\kappa)$. As discussed in \FGPboth, this multiplicity is also the
maximal multiplicity in the standard  weight diagram of the
finite dimensional representation of $sl(3)$ of highest  weight
$3\zl+w(\br)-\br$.

\newsec{Quotient space method }

In this section we shall treat the decoupling problem and the
appearance of finite dimensional solution spaces in an
alternative way.  This method is purely algebraic, and first
appeared in conformal field theory in \FF\ where Feigin and Fuchs
used it to study the fusion in Virasoro minimal models. Since
then it has been applied to other models \GW, and developed
extensively in one direction by Zhu
\Zhu, but we shall stick to the spirit of \FF.

To explain how this method works we shall first reconsider the
case of $sl(2)$ because $sl(2)$ has fewer generators than
$sl(3)$, and hence the expressions are simpler.  This case has
already been treated by Feigin and Malikov \FM\ and by Dong et al
\DLM, with apparently contradictory results, and we think it will
be helpful to explain how the various different results fit
together (n.b. the $\sh(2)$ calculations here are essentially all
contained in \FM\ and \DLM.)

The algebraic method relies on the observation that the inner product
of a highest weight state, a primary field corresponding to the vertex
operator of a highest weight state, and an arbitrary state in a
highest weight representation, satisfies various identities. 

To express these identities, we need some notation. The
generators of $\sh(2)$ are denoted by $ e_m $, $f_m$ and $h_m$
and have commutation relations
\eqna\sltwocomms
$$\eqalignno{
   \comm{h_m}{h_n} 
&= \phantom{-}2\,k\,m\,\delta_{m+n}
\;, 
\cr
   \comm{h_m}{e_n} 
&= \phantom{-}2\,e_{m+n}
\;,
\cr
   \comm{h_m}{f_n} 
&= -2\,f_{m+n}
\;,
\cr
   \comm{e_m}{f_n} 
&= \phantom{-2\,}k\,m\,\delta_{m+n} + h_{m+n}
\;.
& \sltwocomms {}
}$$
We denote highest weight Verma modules with highest weight $\vec
r$ by $M_r$ and the corresponding irreducible module by $L_r$,
where the states $\vec r$ and $\cev{r}$ satisfy
\eqna\sltwohwts
$$\eqalignno{
   f_m \, \vec r 
&= 0
\;,\;\;
   m > 0
\;,\;\;\;\;
   e_m \, \vec r 
 = 0
\;,\;\; 
   m \geq 0
\;,\;\;\;\;
   h_m\,\vec r
 = r\,\delta_{m0}\,\vec r
\;,\;\;
   m \geq 0
\;,
\cr
   \cev{r}\, e_m \, 
&= 0
\;,\;\;
   m < 0
\;,\;\;\;\;
   \cev{r} \, f_m 
 = 0
\;,\;\; 
   m \leq 0
\;,\;\;\;\;
   \cev{r} \, h_m
 = r\,\delta_{m0}\,\cev{r}
\;,\;\;
   m \leq 0
\;.
 & \sltwohwts{}
}$$
We are interested in working out the number of independent couplings
of the form
\eqn\tpfs{
  \cev {\phi_\alpha} \, \phi_\beta(z)\, \vec\psi
\;,
}
where $\vec\psi$ is some general state in a highest-weight
representation of $\sh(2)$; $\cev{\phi_\beta}$ is annihilated by
all $T_m$, $m<0$ (and hence is in some representation of the
`horizontal' $sl(2)$ subalgebra generated by $T \equiv T_0$).
Finally, the fields $\phi_\beta(z)$ transform as
\eqnn\phibeta
$$\eqalignno{
  \comm{ \phi_\beta(z) }{ T_m }
&= z^m\,D^{\vphantom{\phi}}_{\beta\gamma}(T)\,\phi_\gamma(z)
\;,
& \phibeta 
}
$$
where $D(T)$ is some representation of $sl(2)$.
A consequence of this definition is that
\eqn\sltwomore{
  \comm{T_m - z\,T_{m-1}}{\phi_\beta(z)}
= 0
  \;,
}
so that for any generator $T$
\eqn\quotientA{
  \cev {\phi_\alpha} \, \phi_{\beta}(z)\, (T_m - z^{m-1}T_{-1})
= 0
\;,\;\;\;\; m \leq -2
}
Consequently (taking $z{\,=\,}1$) for any state $\vec\psi$, the
state 
$$
  ( T_m - T_{-1} ) \, \vec\psi\,,  \qquad m \leq -2
\;,
$$
has zero inner product with $\cev
{\phi_\alpha}\,\phi_{\beta}(1)$.  This means that to find the
space of independent couplings from an irreducible highest weight
representation $L_{r}$ of $\sh(2)$ to states of the form $\cev
{\phi_\alpha}\, \phi_{\beta}(z)$ one need only consider the
quotient space 
\eqn\CAdef{
  \CA(L_{r})
= L_{r} \big/ < (T_m - T_{-1}) \,L_{r} >
\;.
}

Zhu's algebra $\CA(L_0)$ is exactly the space defined in equation
\CAdef. He showed that the space $\CA(L_0)$ can itself be given
an algebra structure, and that the space of irreducible
representations of the vertex algebra $L_0$ are in one-to-one
correspondence with the irreducible representations of the
algebra $\CA(L_0)$.  The main problem with trying to follow Zhu's
analysis for admissible but non-integrable representations of
$\sh(2)$ is that typically $\CA(L_r)$ is infinite-dimensional for
any $r$, and the analysis of this space correspondingly harder
than for the integrable case (although it may be carried through
-- see e.g. \AM)

However, in the spirit of \FF, it is not necessary to consider
the full space $\CA(L_r)$ to find the allowed fusions with $L_r$.
For the integrable case $k$ one can use the Ward identities to
express the general three-point function
$$
  \cev {\phi_\alpha} \, \phi_\beta(z)\, \vec\psi
\;,
\eqno\tpfs
$$
in terms of some three point function 
\eqn\tpfA{
  \cev{r''}\; \phi_{r'}(x;z)\; \vec{\psi}
\;,
}
where  $\cev{r''}$ is a highest weight state of $\go$ and 
 $\phi_{r'}(x;z)$ is a highest weight state for some Borel subalgebra
 $\bar\gb(x)$ of $\bgo$ parametrised by the coordinate $x$ on the flag
 manifold. 

 For the integrable case the choice of $x$ is not important for the
 reason that $\phi_{r'}(x;z)$ can be expanded as a polynomial in $x$.
 The leading coefficient $\phi_{r'}(0;z)$ then turns out to be a
 highest weight for $\gb \equiv \gb(0)$ and the space of couplings
 turns out to be independent of $x$. 

 However, for an infinite-dimensional representation of $sl(2)$,
 assuming that $\phi_{r'}(x;z)$ can be expanded as an integer power
 series in $x$  requires that $\phi_{r'}(0;z)$ is a highest
 weight for $\gb(0)$ and leads exactly to the
 restricted fusion rules found by BF. To find the full set of fusion
 rules one must accept that one cannot necessarily expand
 $\phi_{r'}(x;z)$ about $x=0$ or $x=\infty$
 and one must take \tpfA\ as a starting point.  
 We shall see that different (non-generic) choices of the Borel
 subalgebra $\bar\gb(x)$ of $\bgo$ used to define $\phi_{r'}(x;z)$ may
 lead for nonintegral $r'$ to different (degenerate) results for the
 space of fusions. 

Given our `standard' splitting into $e_m,f_m$ and $h_m$, 
as for $sl(3)$, with $\bar \gb$ the standard Borel subalgebra of
$sl(2)$ generated by $e_0$ and $h_0$, we can require fields to be
highest weight states for  any conjugate subalgebra
\eqn\algconj{
  \hat \gb = U\, \gb\, U^{-1} \;,
}
of our standard raising operators by some constant
group element $U$. If we consider conjugation by
\eqn\sltwoconj{
  U = \exp(\, x f_0\,)\, \;,
}
for which
\eqn\conjef{
\eqalign{
  \hat e_m
&= \exp(\, x f_0\,)\, e_m\,\exp(\,-xf_0\,)
= e_m - x h_m - x^2 f_m \;,
\cr
  \hat h_m
&= \exp(\, x f_0\,)\, h_m\,\exp(\,-xf_0\,)= h_m + 2 x f_m \;,
}}
then the fields $\phi_{r'}(x;z)$ being highest weight states for these
generators implies 
\eqnn\sltwogenb
$$\eqalignno{
   \comm{e_m - x h_m - x^2 f_m }{\phi_{r'}(x;z)} 
&= 0
\;,
\cr
   \comm{h_m + 2x f_m}{\phi_{r'}(x;z)} 
&= z^m\, {r'}  \, \phi_{r'}(x;z)
\;.
& \sltwogenb 
}$$
If we allow $x$ to take all values including $x{\,=\,}\infty$
(which, suitably interpreted, corresponds to $U{\,=\,}w$, the
Weyl group element of $SL(2)$) then this covers all highest
weight fields.  One can of course find a representation of
$e_m,h_m,f_m$ on the fields $\phi_{r'}(x;z)$ in terms of
differential operators, in a manner analogous to that for
$sl(3)$, such that the relations \sltwomore\ and \sltwogenb\ hold;
however it is not necessary to consider the differential operator
representation explicitly since \sltwomore\ and \sltwogenb\ are
the only relations needed in the algebraic treatment.  So, given
these relations, we find that
\eqna\quotientB $$\eqalignno{
&
 \cev{r''} \, \phi_{r'}(x;1)\, (T_{-m} - T_{-1})
= 0\,, \quad m>1\;,
\cr
& \cev{r''} \, \phi_{r'}(x;1)\, (e_{-1} - x h_{-1} - x^2 f_{-1}) 
= 0
\;,
\cr
&   \cev{r''} \, \phi_{r'}(x;1)\, (f_{0} - f_{-1})
= 0
\;,
\cr
&   \cev{r''} \, \phi_{r'}(x;1)\, (h_{0} - h_{-1})
= {r''}\; \cev{r''}\,\phi_{r'}(x;1)
\;,
\cr
&   \cev{r''} \, \phi_{r'}(x;1)\, (h_{-1} + 2 x f_{-1})
= -{r'}\; \cev{r''}\,\phi_{r'}(x;1)
\;.
& \quotientB {}
}$$
Consequently we define the new quotient space
\eqn\CAltxdef{
  \CA_x^<(L_r)
= L_r \, \big/ \, \CJ_x\,L_r
\;,
}
where $\CJ_x$ is the  linear span of the elements of $\go_-$ 
\eqn\bidea{
   T_{m} - T_{-1}\;, m < -1
\;,\;\;\;\;\;\;
   e_{-1} - x h_{-1} - x^2 f_{-1}
\;,\;\;\;\;\hbox{ and }\;\;
   f_0 - f_{-1}
\;.
}
The space $\CA_x^<(L_r)$ 
then carries an action of $U(1) \oplus U(1)$ with generators
\eqn\Hdefs{
  \Hoo
= h_0 - h_{-1}
\;,\;\;\hbox{ and }\;\;
  \Ho
= -(h_{-1} + 2 x f_{-1})
\;,
}
which take values ${r''}$ and ${r'}$ on
$\cev{r''}\,\phi_{r'}(x;1)$, respectively.
\foot{The generators $\Hoo$ and $\Ho$ which commute
up to an element in $\CJ_x$ are analogues of the 
linear  combinations of Virasoro generators in \FF, 
$L_{0} -2z L_{-1}+ z^2 L_{-2}$ and $L_{-1} - z L_{-2}$ respectively,
while the combinations \bidea\ 
 are  analogues of $L_{-n} -2z L_{-n-1}+ z^2 L_{-n-2}\,,$ $n>0$.}
The space of allowed fusions to the representation $L_r$ is now
the space $\CAL_x(L_r)$ viewed as a $\IC[\Hoo,\Ho]$ module --
that is $(\Hoo,\Ho)$ are restricted to lie on some curves and
points in the $(\Hoo,\Ho)$ plane, possibly with multiplicities.
Although $\Hoo$ and $\Ho$ may not be strictly diagonalisable, we
shall often refer to their allowed values as their `spectrum'; in
the case of $L_r$ admissible it turns out that  $\Hoo$ and $\Ho$
are genuinely diagonalisable, with  $\CAL_x(L_r)$ being a direct
sum of eigenspaces of $\Hoo$ and $\Ho$.  The dimension of the
factor-space of fixed eigenvalues $ \CA_x^<(L_r)^{(r', r'')}$
describes the fusion rule multiplicities,
\eqn\dim{
 {\rm dim} \, 
\CA_x^<(L_r)^{(r', r'')}=N_{r r'}^{r''}\;.
}
A natural spanning set for $L_r / \CJ_x L_r$ are the states
\eqn\spansetA{
  (h_{-1})^m\, (f_0)^n\, \vec{r}
\;.
}
If $x$ is generic, i.e. $x \neq 0,\infty$, then we have 

\eqn\subsA{
  h_{-1} 
= h_0 - \Hoo
\;,\;\;\;\;
  f_{0}  
= (f_0 - f_{-1}) 
+ \frac{1}{2x} ( \Ho - \Hoo - h_0 )
\;.
}
and so an equally good spanning set for $\CAL_x(L_r)$ is 
\eqn\spansetB{
  (\Hoo)^m\; (\Ho)^n\; \vec r
\;.
}
However this is not a good choice if $x{\,=\,}0$ 
(unless $r{\,=\,}0$) or $x{\,=\,}\infty$ (unless $r{\,=\,}k$).  If
$x{\,=\,}0$, then we have $\hat e_{-1}{\,=\,}e_{-1}$, and more
importantly, $h_{-1}{\,=\,}{-}\Ho$ and $\Hoo{-}\Ho{\,=\,} h_0$, so
that $\Hoo$ and $\Ho$ are simultaneously diagonalised on the weight
spaces, and we can take as a spanning set of $\CAL_0(L_r)$
\eqn\spansetC{
   (\Ho)^m\; ( f_0 )^n\; \vec r \;.
}
If, conversely, $x{\,=\,}\infty$, then we find $f_0$ should be
included in $\CI$, that $h_{-1}{\,=\,}{-}\Ho$,
$\Hoo{\,+\,}\Ho{\,=\,}h_0$, so that $\Hoo$ and $\Ho$ are again
diagonalised on the weight spaces, but that $e_{-1}$ in
unconstrained, so that we can take as a spanning set of
$\CAL_0(L_r)$
\eqn\spansetCb{
   (\Ho)^m\; ( e_{-1} )^n\; \vec r
\;.
}

If $M_r$ is irreducible then in each case the states \spansetB,
\spansetC\ and \spansetCb\ are linearly independent, but if $M_r$
contains null vectors, there will be linear relations among these
states. If $x$ is generic, $\CAL_x(L_r)$ naturally takes the form
of a simple quotient of the polynomial ring
$\IC[\,\Hoo\,,\Ho\,]$, since singular vectors in the Verma module
$M_r$ lead to polynomial constraints in $\CAL_x(M_r)$.  However
in the other two cases the structures of $\CAL_0(L_r)$ and
$\CAL_\infty(L_r)$ may be more complicated.

Physically, taking $x\,{=}\,0$ or $x\,{=}\,\infty$ puts strong
constraints on the allowed fusions.  For representations for
which $r$, $r'$ or $r''$ are not  non-negative integers,
$$  
  \cev{r''}\; \phi_{r'}(x;z)\; \vec{r} \;,
$$
may have singular expansions around $x{\,=\,}0$ and
$x{\,=\,}\infty$.  One would hope that the algebraic method would
only find the fusions for which the three point functions are
regular, that the dimension of $\CAL_0(L_r)$ and
$\CAL_\infty(L_r)$ would be smaller than the generic result
$\CAL_x(L_r)$, excluding precisely those fusions for which the
three-point function does not have a power-series expansion at
$x=0$ and $x=\infty$ respectively.  This is exactly what has been
found.

Feigin and Malikov calculated $\CAL_x(L_r)$ for the generic
values $x\!=\!1$ and found the `nice' fusion rules of Awata \&
Yamada \AY; The fusion rules were also investigated in \DLM\
through the construction of a space `$\CA(L)$' -- which is
nothing but $\CAL_x(L_r)$ with $x\!=\!0$ -- and instead of the
fusion rules of \AY, they found the fusion rules of Bernard \&
Felder \BF; these fusion rules are a `subset' of the rules of
\AY\ in the sense that all couplings allowed by \BF\ are allowed
by \AY, but not all the \AY\ couplings are allowed by \BF.

Taking $x\!=0\!\,$ is not `wrong' in a mathematical  sense, but
rather the mathematics gives the right answer to what may be the
wrong physical question.

We illustrate these points in the simplest non-trivial cases.  If
we define $\kappa\,{=}\,k{+}2$, then the simplest representations
containing singular vectors which are not of the form $f_0^m \vec
r$ or $e_{-1}^n   \vec r$ are \hbox{$r\,{=}\,{-}\kappa$} and
\hbox{$r\,{=}\,2\kappa{-}2$}, with singular vectors
\eqnn\svone
\eqnn\svtwo
$$\eqalignno{
   \vec{1}
&= \Big(\;\,\,
   f_0\,f_0\,e_{-1}
\,\;+\;\,
   (1{-}\kappa)\,f_0\,h_{-1}
\,\;+\;\,
   \kappa(1{-}\kappa)\,f_{-1}
   \;\Big)
   \vec{-\kappa}
\;,
&\svone 
\cr
\noalign{\noindent
and
}
   \vec{2}
&= \Big(\;
   e_{-1}\,e_{-1}\,f_0
\;-\;
   (1{-}\kappa)\,e_{-1}\,h_{-1}
\;+\;
   \kappa(1{-}\kappa)\,e_{-2}
   \,\Big)
   \vec{2\kappa{-}2}
\;,
&\svtwo
}$$
respectively.  These are the two simplest non-trivial
representations in the set of `pre-admissible' representations,
of spins
\eqn\preadmA{
  r \in
\{\; 
  n' - n \kappa
\;,\;\; n,n' = 0,1,2,\ldots
\;;\quad
  -n' + (n+1)\kappa
\;,\;\; n,n' = 0,1,2,\ldots
\;\}
\;.
}
It is expected that these representations form a closed
subalgebra for generic $\kappa \not = \IQ$.

The admissible representations have $\kappa$ certain rational
numbers and spins $r$ a subset of the `pre-admissible'
representations,
\eqn\admA{
  \kappa 
= p'/p 
\;,\;\;\;\;
  r \in \{ n' - n\kappa = -(p'-n') + (p-n)\kappa
           \;,\;\;
           0\leq n'\leq p'-2\;,\;\;
           0\leq n \leq p -1
          \}
\;.
}
The simplest non-trivial admissible representation is at level
$k{\,=\,}{-}4/3$, $\kappa{\,=\,}2/3$, of spin
$r{\,=\,}{-}\kappa{\,=\,}2\kappa{-}2{\,=\,}{-}2/3 $; hence both
\svone\ and \svtwo\ are in $M_{-\kappa}$ in this case, and are
the linearly independent generators of the maximal submodule of
$M_{-\kappa}$. For this level, there are only three admissible
representations, $r \in\{0,-\kappa,-2\kappa\}=\{0,-2/3,-4/3\}$,
and we again expect the fusion algebra to be closed on this set.

\subsec{The case of $x$ generic}

Let us first consider the case $x\neq 0,\infty$. 
\foot{In other words, as stressed in   \FM, one assigns distinct
Borel subalgebras   to the three fields in the correlator, or,
choosing $x=1$, we have $\bar{\gb}_0= e\oplus h$,
$\bar{\gb}_{\infty}= f\oplus h$, $\bar{\gb}_1= (e-h-f)\oplus
(h+2f)$. }
Using equations \subsA\ it is straightforward to show that in
$\CAL_x(L_r)$ the singular vectors $\vec 1$ and $\vec 2$ are
equivalent to
\eqnn\svoneR
\eqnn\svtwoR
$$\eqalignno{
  \vec 1
&\cong
  -\frac{1}{8x}
  \Big( \Ho - \Hoo - \kappa       \Big)
  \Big( \Ho - \Hoo + \kappa       \Big)
  \Big( \Ho + \Hoo - (\kappa{-}2) \Big)
  \vec{-\kappa}
\qquad~
&\svoneR
\cr
  \vec 2
&\cong
  -\,\frac{x}{8}\,
  \Big( \Ho + \Hoo + 2(1{-}\kappa)\Big)
  \Big( \Ho + \Hoo + 2            \Big)
  \Big( \Ho - \Hoo                \Big)
  \vec{2\kappa{-}2}
\qquad~
&\svtwoR
}$$
This means that these singular vectors each lead to a single
polynomial constraint between $\Ho$ and $\Hoo$.  If $\Ho$ is in
the `pre-admissible set', then so is $\Hoo$, except for the cases
of $\Ho$ being on the `edge' of the `pre-admissible set', that is
one of $n$ or $n'$ being zero, in which case the simple
constraints arising from the null-vectors in the representation
$\Ho$ must also be taken into account. In each case, the fusion
of these two representations with representations ${r'}$ in the
pre-admissible set yields fields ${r''}$ in the pre-admissible
set.
 
The values of $(\Hoo,\Ho)$ which are allowed to couple to the
representation $\vec 1$ are shown in figure 1.

In the case that $\kappa$ takes the admissible value $\kappa=2/3$, we
then have ${-}\kappa{\,=\,}2\kappa{-}2$, and so $\Ho$ and $\Hoo$  must
satisfy the two simultaneous equations 
\eqna\simeqs
$$\eqalignno{
  \Big( \Ho - \Hoo - \fracs 23       \Big)
  \Big( \Ho - \Hoo + \fracs 23       \Big)
  \Big( \Ho + \Hoo + \fracs 43       \Big)
&= 0
&\simeqs a
\cr
  \Big( \Ho + \Hoo + \fracs 23     \Big)
  \Big( \Ho + \Hoo + 2            \Big)
  \Big( \Ho - \Hoo                \Big)
&= 0
&\simeqs b
}$$  
The solutions to these simultaneous equations are 
\eqn\simsols{
   (\Hoo,\Ho)
\;\in\;
  \Big\{\;
    ( -{\fracs{2}{3}},-{\fracs{4}{3}}) \;,\;
    ( -{\fracs{4}{3}},-{\fracs{2}{3}}) \;,\;
    ( -{\fracs{2}{3}},-{\fracs{2}{3}}) \;,\;
    ( 0              ,-{\fracs{2}{3}}) \;,\;
    ( -{\fracs{2}{3}},0) 
  \;\Big\}
\;,
}
all of which lie in the set of admissible representations with no
further constraints necessary.  This corresponds to a fusion
matrix for the fundamental field $f{\,=\,}({-}\kappa)$ of the
form
\eqn\fusf{
  \CN_f
= \pmatrix{ 0 & 1 & 0 \cr 1 & 1 & 1 \cr 0 & 1 & 0 }
\;.
}

\subsec{The case $x {\,=\,} 0$}

In this case we have to consider the constraints arising from the
vanishing of all \hbox{$f_0$--descendents} of the singular vectors.
Taking $\vec 1$ first, it is easy to show that the only relations
among the spanning set \spansetC\ arise from
\eqnn\svoneRb
$$\eqalignno{
   (f_0)^p\, \vec 1 
&= \Big(\,
   e_{-1}\,(f_0)^{2}
-  (\kappa+p+1)\,h_{-1}\,f_0
-  (\kappa+p+1)(\kappa+p)\,f_{-1}
   \;\Big)\,(f_0)^p\,
   \vec{-\kappa}
\cr
  (f_0)^p\,\vec 1 
& \cong
  (\kappa+p+1)\,
  \Big(\, \Ho -\kappa -  p  \,\Big) \,
  (f_0)^{p+1}\,\vec{-\kappa}
\;.
& \svoneRb 
}
$$
This makes for quite a complicated structure. For generic $\kappa$,
\eqn\structA{
  \CAL_0(L_{-\kappa})
= \IC[\Ho]\,\vec{-\kappa}
  \oplus
  \Big(
  \oplus_{p>0} \IC\,(f_0)^p\,\vec{-\kappa}
  \Big)
\;.
}
On the first summand $\Hoo$ and $\Ho$ are only restricted by
$\Hoo - \Ho {\,=\,} -\kappa $, but on each of the discrete
eigenvectors $(f_0)^p\vec{-\kappa}$ they take values $\{ \Hoo
{\,=\,} -p-1\;,\;\;  \Ho {\,=\,} \kappa+p-1\}$.  This is exactly
the subset of the spectrum at generic $x$ which satisfies $\Ho
-\Hoo - \kappa {=}0,2,4,\ldots$, and is also shown on figure 1.
The algebraic method exactly reproduces the physically intuitive
result - putting $x{\,=\,}0$, one restricts to the subset for
which the three-point functions have a regular expansion around
$x{\,=\,}0$.  (This is essentially the same restriction as that
imposed by \BF.)

Taking now the case of $\vec 2$, we have
\eqnn\svtwoRb
$$\eqalignno{
  (f_0)^p\,\vec 2 
& \sim 
  -p(\kappa{-}p)
  \Big( \Ho -(p-1)  \Big) 
  \Big( \Ho +\kappa-p+1 \Big)\, 
  (f_0)^{p-1}\, 
  \vec{2\kappa{-}2}
\;,~~~~~~~~~
& \svtwoRb 
}
$$
so that 
\eqn\structB{
  \CAL_0(L_{2\kappa{-}2})
= \oplus_{p\geq 0} 
  \Big(
  \IC\, (f_0)^p\,\vec{2\kappa{-}2}
  \oplus
  \IC\, \Ho\,(f_0)^p\,\vec{2\kappa{-}2}
  \Big)
\;.
}
On each of the two dimensional spaces in the sum in \structB,
$\Hoo$ and $\Ho$ are diagonalisable with joint eigenvalues $(
2\kappa {\,-\,} p {\,-\,}2\,,\,p)$ and $(  \kappa {\,-\,} p
{\,-\,}2 \,,\, -\kappa {\,+\,} p )$.  This is now the subset of
the spectrum at generic $x$ which satisfies $\Ho {\,-\,} \Hoo
{\,+\,} 2 \kappa{-}2 {\,=\,}0,2,4\ldots$

In the admissible case $\kappa {\,=\,} 2/3$, both \svoneRb\ and
\svtwoRb\ must be set to zero in $\CAL_0(L_{-2/3})$.  This
reduces the space to two dimensions,
\eqn\spansetD{
  \IC\,\vec{-2/3}\;
  \oplus\; 
  \IC\,\Ho\,\vec{-2/3}
\;,
}
on which the eigenvalues are 
\eqn\simsolsB{
   (\Hoo,\Ho)
\;\in\;
  \Big\{\;
  ( -{\fracs{4}{3}},-{\fracs{2}{3}}) \;,\;
  ( -{\fracs{2}{3}},0) 
  \;\Big\}
\;,
}
a subset of \simsols, agreeing with the results of Bernard \& Felder.
This would correspond to a fusion matrix for the fundamental field of
the form 
\eqn\fusf{
  \CN_f
= \pmatrix{ 0 & 1 & 0 \cr 0 & 0 & 1 \cr 0 & 0 & 0 }
\;,
}
which is rather degenerate  -- for example, it is nilpotent, and
there is no conjugate field $f^*$ such that the identity would
appear in the fusion of $f$ and $f^*$.

\subsec{The case $x {\,=\,} \infty$}

The case of $x{\,=\,}\infty$ is analogous to that of $x{\,=\,}0$, but
since the representation $\cev{r''}$ can also be thought of as having
$x=\infty$, the results are now symmetric in $\Ho$ and $\Hoo$.
In each case we find that $\CAL_\infty(L_{r})$ consists 
of the subset of the spectrum for $x$ generic
satisfying \hbox{$\Ho+\Hoo{\,-\,}{r}{\,=\,}0,2,4\ldots$}.

For $\CAL_\infty(L_{-\kappa})$ this is now a set of discrete points
while (as shown in figure 1) and for
$\CAL_\infty(L_{2\kappa-2})$ this a line and a set of points. 

In the rational case $\kappa=2/3$, the $\CAL_\infty(L_{-\kappa})$ is
again two-dimensional, the spectrum this time consisting of
$(-\kappa,0)$ and $(0,-\kappa)$, leading to a `fusion' matrix 
\eqn\fusfB{
  \CN_f = \pmatrix{ 0 & 1 & 0 \cr 1 & 0 & 0 \cr 0 & 0 & 0 } \;.
}
Although the matrix \fusfB\ is symmetric (unlike \fusf) 
and thus there is  a conjugate $f^* =f$, it is however
decomposable, i.e.\ as  for $x=0$  the result for
$x=\infty$ does not
satisfy the usual requirements for a  fusion algebra.
\foot{ This case appears implicitly  in \BF, 
 where it is mentioned that both lowest and  highest weight
chiral vertex operators are
needed to have in particular  nonvanishing $2$--point functions;
the $4$--point blocks construction in \D{} provides  explicit
examples of ``mixed'' correlators.  They can be interpreted as
proper limits with coordinates $x_i$ taken at $0$ or $\infty$ of
a subset of the generic conformal blocks in \FGPP, \PRY.
}

\vskip .2cm
\overfullrule0pt
$$
\epsfxsize=.95\hsize
\eqalign{
&\epsfbox{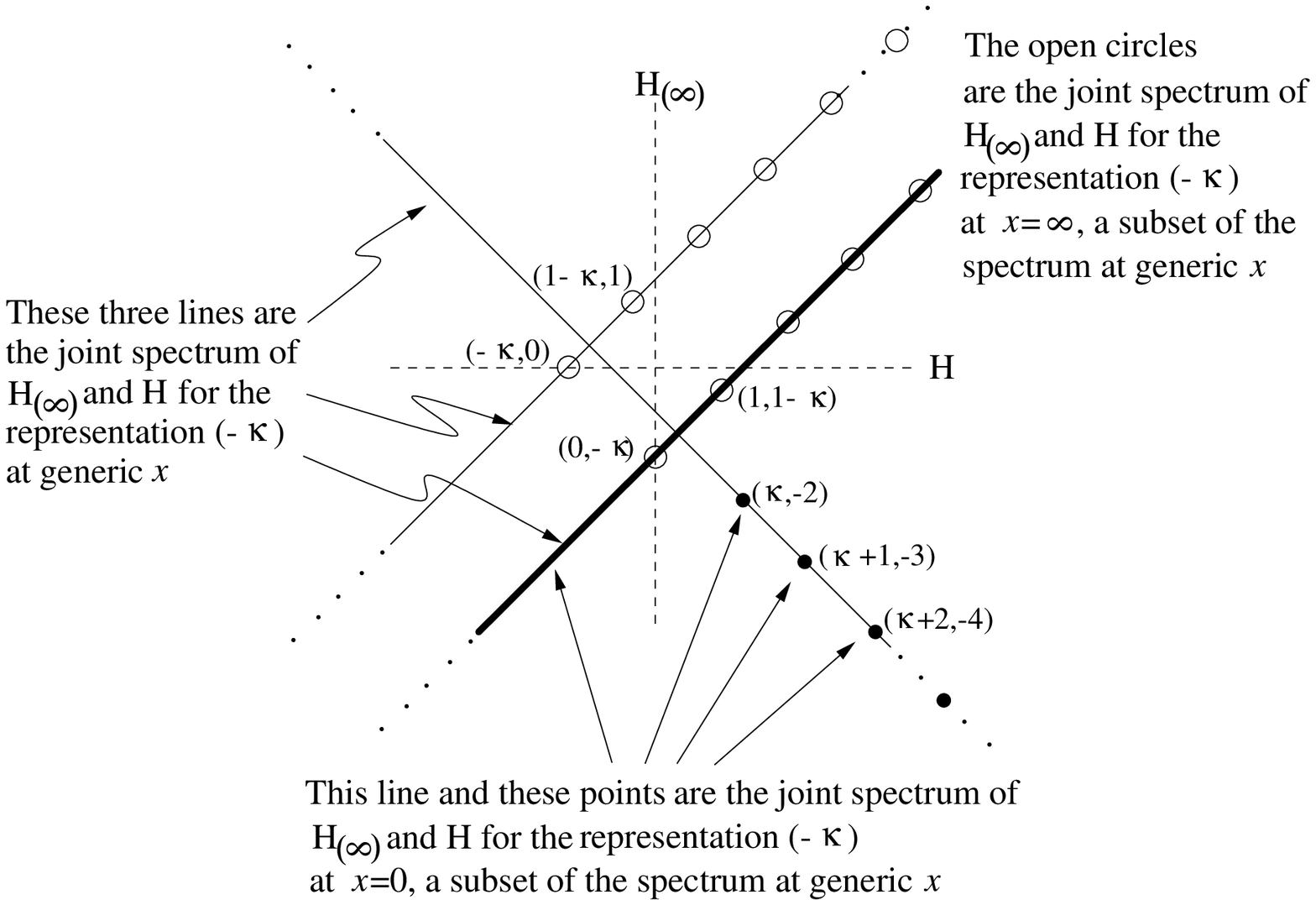}
\cr\cr
&\hbox{\phantom{Figure 1:~~~~~~~~~~}
\vbox{\openup-2\jot
\noindent
\llap{Figure 1: }
The `spectrum' of $\Hoo$ and $\Ho$ on
$\CAL_x(L_{-\kappa})$ for
generic $\kappa$,
\hfill\break\noindent
 showing the results for generic $x$, $x{\,=\,}0$ and $x\,{=}\,\infty$.
\vfill}}
}
$$
\vskip .2cm

\subsec{The case of $sl(3)$}

All the preceding discussion passes over to the case of $\sh(3)$,
with some minor modifications and some unfortunate complications.
We recall that the fields $\Phi_\mu(X)$ we introduced earlier are 
highest weight primary fields for the conjugate algebra generators,
satisfying  
$$\eqalignno{
  \comm{T_m - T_{m-1}}{\Phi_\mu(X)}
&= 0
\;,
\cr
  \comm{\hat e^i_m}{\Phi_\mu(X)}
&= 0
\;,
\cr
  \comm{\hat h^i_m}{\Phi_\mu(X)}
&= \mu(h^i)\,\Phi_\mu(X)
\;.
}$$
We are interested in the space of three-point functions 
\eqn\tpfD{
  \cev\gamma\, \Phi_\mu(X)\, \vec\psi \;,
}
for some highest weight state $\cev\gamma$ and $\vec\psi$ an
arbitrary state in some irreducible highest weight representation
$L_\nu$.  By analogy to the case of $\sh(2)$ we shall define
$\CAL_X(L_\nu)$ to be the quotient
\eqn\calB{
  \CAL_X(L_\nu)
= L_\nu \, \Big/ \, \CI L_\nu
\;,
}
where for finite $X$ $\CI$ is the  linear span of the elements
\eqn\ida{
  T_{-n}   {-} T_{-1}\;,\;\; n>2\;,\;\;\;\;\;\;\;\;
  f_{-1}^j {-} f_0^j \;,\;\;       \;\;\;\;\;\;\;\;
  \hat{e}_{-1}^j   \;.
}
As before, we define 
\eqn\hdefs{
  \Hoo^i
= h^i_0 - h^i_{-1}
\;,\;\;\;\;
  \Ho^i
= - \hat h^i_{-1}
\;,
}
which commute on $\CAL_X(L_\nu)$, and the space of fusions to
$L_{\nu}$ is given by the `joint spectrum' on $\CAL_X(L_\nu)$ of these
operators.  

{}From \calB\ and definition of $\Ho^i$, it is clear that
for finite $X$, $\CAL_X(M_\nu)$ 
is spanned by the states 
\eqn\spansetE{
   (\Ho^1)^{b_1} \, 
   (\Ho^2)^{b_2} \, 
   (f^1_0)^{a_1}    \,
   (f^2_0)^{a_2}    \,
   (f^3_0)^{a_3}    \, 
   \vec\nu
\,,  \quad a_i\,, \, b_j=0,1, \dots 
\,
}
For generic $X$ it is further possible to show that the space 
$\CAL_X(M_\nu)$ 
is also spanned by the states
\eqn\spansetF{
   (\Hoo^1)^{a_1} \, 
   (\Hoo^2)^{a_2} \, 
   (\Ho^1 )^{a_3}    \,
   (\Ho^2 )^{a_4}    \,
   (f^3_0)^{a_5}    \, 
   \vec\nu
\,,  \quad a_i=0,1, \dots 
\,. 
}
The values of $X$ on which this inversion is not possible are
similar to the points $x=0,\infty$ for $\sh(2)$, but 
are no longer just points but subspaces.

The main technical difficulty in computing 
$\CAL_X(L_\nu)$, even for generic $X$,
is that since we cannot assume $a_5{\,=\,0}$ in \spansetF,
$\CAL_X(M_\nu)$  is not just $\IC[ \Hoo^i,\Ho^i]$,
singular vectors in $M_\nu$ do not just lead to polynomial constraints
on $\Hoo^i$ and $\Ho^i$, and 
$\CAL_X(L_\nu)$ is not just
automatically a quotient of  $\IC[ \Hoo^i, \Ho^i ]$.
This fact is intimately connected with the possible existence 
of Verlinde fusion numbers greater than 1, as we shall see. Rather,
(as for the case $x=0$ in $\sh(2)$) we must also consider descendents
of the highest-weight singular vectors, and the explicit construction
of the space  $\CAL_X(L_\nu)$  is rather messy.

We shall not attempt any further general discussion, but simply
outline the results of an explicit calculation in the simplest
`pre-admissible' representation which has fusions with non-trivial
multiplicities.

\subsec{The representation $h$} 

The representation `$h$' has weight
$\bar\rho(\kappa - 2)$ 
and we shall denote the
highest weight state by $\vec h$.
The irreducible representation $L_h$ is the quotient of the Verma
module $M_h$ by its maximal submodule which is generated by the two
independent singular vectors 
\eqn\singvecs{
\eqalign{
   \vec 1 
&= \Big(\,
    e^3_{-1}\,f^2_0 - (2-\kappa)\,e^1_{-1} \, 
   \Big)\,\vec h
\;,
\cr
   \vec 2 
&= \Big(\,
    e^3_{-1}\,f^1_0 + (2-\kappa)\,e^2_{-1} \, 
   \Big)\,\vec h
\;.
}}
Using the relations generated by \ida, the singular vectors
\singvecs\ are equivalent in $M_h / \CJ M_h $ to
\eqna\singvecsB
$$
\eqalignno{
   \vec 1 
&\cong
   \Big(\,
   \fracs 43 x \,  \Ho^1
+  \fracs 43 x^2 \,f^1_0
+  ( \Ho^2 (x y - z)
   - \Ho^1 z
   - \fracs 43 (x y - z))\, f^2_0
+ \fracs 43 x^2 y f^3_0
\cr&\phantom{\sim\Big(\,}
- x z f^1_0 f^2_0
+ y(x y - z) f^2_0 f^2_0
+ ( x^2 y^2 + z(x y - z)) f^3_0 f^2_0
   \,\Big)\, \vec h
& \singvecsB a
\cr
   \vec 2 
&\cong
   \Big(\!
-  \fracs 43 y \,  \Ho^2
-  \fracs 43 y^2 \,f^2_0
+  ( \Ho^2 (x y - z)
   - \Ho^1 z
   + \fracs 43 z)\, f^1_0
+ ( \fracs 43 x y^2 
  + y(x y - z) )f^3_0
\cr&\phantom{\sim\Big(\,}
- x z f^1_0 f^1_0
+ y(x y - z) f^1_0 f^2_0
- ( x^2 y^2 - z(x y - z)) f^3_0 f^1_0
   \,\Big)\, \vec h
\;.
& \singvecsB b
}$$
As is obvious, just considering the constraints in $\CAL_X(L_h)$
from these two singular vectors does not lead to any restrictions
on $\Hoo^i$ and $\Ho^i$.  At the moment we do not have an
analytic method to analyse $\CAL_X(L_h)$, but using Mathematica
we have investigated explicitly the constraints arising from many
(up to 56) $f^i_0$--descendents of the singular vectors $\vec 1$
and $\vec 2$, and have found that $\CAL_X(L_h)$ is (at largest) a
direct sum
\eqn\fivedimA{
  \CAL_X(L_h)
= \oplus_{i=1}^5 \; \IC[\Ho^1,\Ho^2]\cdot v_i
\;,
}
where one choice of $v_i$ is:
\eqn\fivedimB{
\eqalign{
   v_1
&=  (\Hoo^1 {-} \Ho^1)
    (\Hoo^1 {-} \Ho^1 {-} \Ho^2 {-} 1) 
    (\Hoo^1 {+} \Ho^2 {+} 2{-} \kappa) 
    \,\vec h
\cr
   v_2
&=  (\Hoo^1 {-} \Ho^1)
    (\Hoo^1 {-} \Ho^1 {-} \Ho^2 {-} 1) 
    (\Hoo^1 {+} \Ho^1 {+} 2) 
    \,\vec h
\cr
   v_3
&=  (\Hoo^1 {-} \Ho^1)
    (\Hoo^1 {+} \Ho^1 {+} 2) 
    (\Hoo^1 {+} \Ho^2 {+} 2{-} \kappa) 
    \,\vec h
\cr
   v_4
&=  (\Hoo^1 {+} \Ho^1 {+} 2) 
    (\Hoo^1 {-} \Ho^1 {-} \Ho^2 {-} 1) 
    (\Hoo^1 {+} \Ho^2 {+} 2{-} \kappa) 
    \,\vec h
\cr
   v_5
&=  (\Hoo^1 {+} \Ho^1 {+} 2) 
    (\Hoo^1 {-} \Ho^1 {-} \Ho^2 {-} 1) 
    (\Hoo^1 {+} \Ho^2 {+} 2{-} \kappa) 
    \,f^3_0\,\vec h
}}
Of course this representation of the $v_i$ is not unique, but it
has one advantage in that it is independent of $X$.  On each of
these vectors the actions of $\Hoo^1$ and $\Hoo^2$ are given in
terms of $\Ho^1$ and $\Ho^2$ by
\eqn\fivedimC{
(\Hoo^1,\Hoo^2) = 
\cases{
\phantom{v_4\;,\;}
  v_1:\quad
& $(- \Ho^1-2  \,,\,   \Ho^1 + \Ho^2 + 1\,)\;,\;\;$
\cr
\phantom{v_4\;,\;}
  v_2:\quad
& $(-\Ho^2-2 + \kappa \,,\, -\Ho^1-2 + \kappa\,)\;,\;\;$
\cr
\phantom{v_4\;,\;}
  v_3:\quad
& $(\;\Ho^1 +1 + \Ho^2 \,,\, - \Ho^2 -2 \,)\;,\;\; $
\cr
  \!v_4\;,\;v_5:\quad
& $(\;\Ho^1\,,\, \Ho^2\,)\;.$
}}
This is exactly the same space of allowed fusions as we found in
section 3, with the same multiplicities.  Note that the
representation of $v_4$ presented in \fivedimB\ uses $f^3_0$ --
this is essential to provide the multiplicity two for the
eigenvalues $\Hoo^i=\Ho^i$.  

 Viewed as a representation of $\IC[\Ho^1,\Ho^2]$, $\CAL_X(L_h)$ is
 five-dimensional. We can also choose to fix the value of $\Ho^i$ to
 some generic weight $\mu$ by extending the set \ida\ generating the
 right ideal  by the combinations $\Ho^i- \mu(h^i)$. This leads to the
 smaller space $\CAL_X(L_h)^{\mu}$ which is genuinely a
 five-dimensional vector space over $\IC$.
 There is no canonical choice for the representatives of a basis of 
 $\CAL_X(L_h)^{\mu}$, but we note that one choice of vectors spanning
 the space is the set 
 $$
       \{
 \;    \vec h
 \;,\; f^1 f^2 \vec h
 \;,\; f^2 f^1 \vec h
 \;,\; f^1 f^3 \vec h
 \;,\; f^2 f^3 \vec h
 \;    \}
 \;.
 $$
 Although the action of $h^i_0$ on this space has no immediate
 connection with the actions of $\Hoo^i$ or $\Ho^i$, this choice does
 exhibit clearly a connection with the subset of the weight diagram of
 the $sl(3)$ adjoint representation of highest weight
 $\bar{\rho}=\iota(w_0)$ -- the image of 
 $h=\overline{w_0\cdot k\Lambda_0}$ under  the map $\iota$ described
 in \FGPboth.  
For $\Ho^i =0$  the set of eigenvalues of $\Hoo^i$ in
\fivedimC\ together with their multiplicities recovers the
generalised weight diagram associated with the representation
$h$. Similarly  for any `preadmissible' weight $\nu$ the
eigenvalues of $\Hoo^i$ in the space $\CAL_X(L_{\nu})^{0}$ define
a generalised weight diagram.  While  for the `pre-integrable'
sub-series of weights in \pre\ this coincides with the standard weight
diagram $\Gamma_{\nu}$ of the finite dimensional representation
of $sl(3)$ of highest weight $\nu$, we expect that   the
generalised weight diagrams introduced in \FGPboth\ will be
reproduced for the sub-series of \pre\ with $\lambda'=0$.

We have also repeated the same analysis for the  
representation $f$  in \fun\
and this time find $\CAL_X(L_f)$ to be a  $\IC[\Ho^1,\Ho^2]$
module as \fivedimA\ with \fivedimB\ replaced by a seven
dimensional space.  Each set of eigenvalues have multiplicity one
in this case, agreeing with the results of \FGPboth.

For non-generic $X$, e.g., 
any of $x,y$ or $z$ being $0$ or $\infty$, or satisfying
$xy-z=0$, the analysis leading to these results breaks down and the
spaces $\CAL_X(L_{\nu})$ are expected to become more complicated. 

It would be nice to find a general method to treat the case of
$\sh(3)$ rather than have to use explicit calculations in each
case, and this is a problem to which we hope to return in the
future.

\bigskip

\newsec{Conclusions}

In this paper we have employed two  methods of dealing with the
null vectors decoupling constraints on some $3$--point $sl(3)$
invariants.  Both lead to the same result and are in full
agreement with the fusion rules determined in \FGPboth.  We have
concentrated here on  ``preadmissible'' representations
characterised by a generic value of the level, the additional
singular vectors arising at rational level values $k+3=p'/p$ can
be  similarly analysed. 

The case of non integral (dominant) highest weights reveals a new
phenomena, namely a dependence of the fusion rules on the
coordinates of the flag manifold.  Taking $x=0$ or $x=\infty$ in
the $\sh(2)$ $3$--point decoupling equations leads to a subset of
the generic fusions of \AY, \FM\ -- the rule in \BF, which gives
three-point functions with regular power expansions around these
points. 
The extension of this analysis to $\sh(3)$, with the fusion rules
in \FGPboth\ corresponding to generic $X$, is possible and
reasonably straightforward, if messy.

We have shown how the approach of Feigin and Malikov produces
the correct results for $sl(3)$ as well as for $sl(2)$.  In \DLM,
Dong et al.~found the same results for the space of fields
$\CAL_0(L_0)=\CAL_x(L_0)$ as Feigin and Malikov, and called this
space a `$\IQ$--graded Zhu's algebra'. This has the calculational
advantage that it is finite dimensional for all admissible
models, whereas Zhu's algebra itself is only finite dimensional
for unitary models, but the disadvantage that it is too ``small''
and does not agree with Zhu's algebra in any of the latter cases
except the trivial case of $k=0$.  There is clearly some point in
extending Zhu's methods to cover general admissible models, but
supposing that \DLM\ is along the right route, it will certainly
be necessary to consider more general spaces $\CAL_X(L_{\nu})$
for arbitrary $\nu$ rather than simply $\CAL_0(L_{\nu})$ as in
\DLM\ to recover the full non-degenerate fusion rules.

The computations here are still not sufficient for a full proof
that the fractional level fusion rule multiplicities in \FGPboth\
are precisely the ones resulting from the solutions of the
singular vectors decoupling equations at generic $X$.  However
they strongly support this expectation in confirming the basic
rules for all ``fundamental'' representations generating the
fusion ring.

The algebraic and differential equation methods are clearly
equivalent, and for the model presented here the differential
equation method is much easier to understand and faster to
analyse.  However, there are algebras, such as the $W_3^{(2)}$
algebra of Bershadsky--Polyakov for which there is no known
action of the algebra on the fields in terms of differential
operators, but for which a naive application of the algebraic
method has so far only produced fusion rules akin to those of
\BF\ for $sl(2)$, with similar problems (nilpotency etc). We
expect that the insight gained from these calculations, and the
way in which degenerate fusion rules can be seen as coming from
an incomplete parametrisation of the space of primary fields,
will lead to progress on such problems.

\bigskip 
\bigskip 
\noindent{\bf Acknowledgements}
\medskip 

We would like to thank F.~Malikov for very helpful discussions at
various times.

This work was supported by an agreement between the
Bulgarian Academy of Sciences and Royal Society, UK.
V.B.P.  acknowledges the hospitality of the Math. Department of
King's College, London, and of the Arnold Sommerfeld Inst.\ f.\
Math.\ Phys.\ of TU Clausthal, as well as partial support of the
Bulgarian National Research Foundation (contract $\Phi-643$).
G.M.T.W. thanks the EPSRC (UK) for an advanced fellowship, 
W.~Eholzer, M.R.~Gaberdiel, P.~Mathieu and M.A.~Walton for useful
discussions.

\bigskip
\bigskip

\listrefs

\bye